\documentclass[aps,showpacs,preprintnumbers,amsmath,amssymb,superscriptaddress,floatfix,nofootinbib]{revtex4-1}
\usepackage{lipsum}
\usepackage{graphicx}
\usepackage{epsfig}
\usepackage{epstopdf}
\usepackage{hyperref}
\usepackage{amsmath}
\usepackage{amsfonts}
\usepackage{amssymb}
\usepackage{caption}
\usepackage{subcaption}
\DeclareMathSizes{12}{12}{12}{12}
\usepackage[section]{placeins}
\usepackage{slashed}
\begin{document}
\title{Coupled channels dynamics in the generation of the $\Omega (2012)$ resonance}
\date{\today}
\author{R. Pavao}
\email{rpavao@ific.uv.es}

\author{E.~Oset}
\affiliation{Departamento de
F\'{\i}sica Te\'orica and IFIC, Centro Mixto Universidad de
Valencia-CSIC Institutos de Investigaci\'on de Paterna, Aptdo.22085,
46071 Valencia, Spain}

\begin{abstract} 
We look into the newly observed $\Omega (2012)$ state from the molecular perspective in which the resonance is generated from the $\bar{K} \Xi^*$, $\eta \Omega$ and $\bar{K} \Xi$ channels. We find that this picture provides a natural explanation of the properties of the $\Omega (2012)$ state. We stress that the molecular nature of the resonance is revealed with a large coupling of the $\Omega (2012)$ to the $\bar{K} \Xi^*$ channel, that can be observed in the $\Omega (2012) \rightarrow \bar{K} \pi \Xi$ decay which is incorporated automatically in our chiral unitary approach via the use of the spectral function of $\Xi^*$ in the evaluation of the $\bar{K} \Xi^*$ loop function.
\end{abstract}

\maketitle
\section{Introduction}

The recent observation of an excited $\Omega$ state, $\Omega^*$, by the Belle collaboration in the $K^- \Xi$ and $K^0_S \Xi^-$ decay channels \cite{Yelton:2018mag} has stirred a new wave of theoretical papers aiming at explaining the nature of the state and its decay channels.

The existence of excited $\Omega$ states, like for any other baryon, is predicted in the quark models by means of the excitations of the quarks \cite{Chao:1980em,Capstick:1986bm,Loring:2001ky,Pervin:2007wa,Faustov:2015eba}. Large $N_c$ considerations \cite{Goity:2003ab}, QCD sum rules \cite{Aliev:2018syi,Aliev:2018yjo}, the Skyrme model \cite{Oh:2007cr} and lattice QCD simulations \cite{Engel:2013ig} have also added information to this subject. Extensions of quark models which would accommodate five quark components  \cite{Yuan:2012zs,An:2013zoa,An:2014lga} lead to more binding than the original ones of Refs.~\cite{Chao:1980em,Capstick:1986bm}. Another extension of the quark model is done in Ref.~\cite{Xiao:2018pwe} using the chiral quark model.

We investigate the state from the molecular point of view with coupled channels and a chiral unitary approach. Work along these lines was done in Refs.~\cite{Kolomeitsev:2003kt,Sarkar:2004jh} and more recently in Ref.~\cite{Si-Qi:2016gmh}. Since the state is close to the $\bar{K} \Xi^*$ threshold ($\Xi^*$ of the $\Delta$ decuplet), the channel $\bar{K} \Xi^*$ is the essential one, but the coupled channels require to consider simultaneously the $\eta \Omega$ channel. This system is, however, peculiar since the $\bar{K} \Xi^* \rightarrow \bar{K} \Xi^*$ interaction with the dominant Weinberg-Tomozawa interaction is zero (and so is the direct $\eta \Omega \rightarrow \eta \Omega$ one). This means that the $\bar{K} \Xi^*$ system by itself does not bind. It is the interaction with the $\eta \Omega$ channel that finally leads to a bound state. This peculiar behavior has, however, an undesired side effect since the predictions of the model are very sensitive to the way the loops are regularized. It suffices to mention that in Ref.~\cite{Kolomeitsev:2003kt} the binding can be obtained around $1950$ MeV, in Ref.~\cite{Sarkar:2004jh} using dimensional regularization with a subtraction constant $a\simeq -2$ a state around $2142$ MeV was found, while the same model with a subtraction constant around $-3.4$ leads to a mass around $1785$ MeV \cite{Si-Qi:2016gmh}. It is clear that the model allows much flexibility, as a consequence of the null diagonal matrix elements of the interaction. The observation of the $\Omega (2012)$ state close to the $\bar{K} \Xi^*$ threshold has provided us with vital information to control the chiral unitary approach. One can use the experimental data to constrain the regulator of the loop functions and then make further predictions to be contrasted with experiment.

After the experimental observation, work along this chiral unitary model was done in Refs.~\cite{Polyakov:2018mow,Valderrama:2018bmv,Lin:2018nqd,Huang:2018wth}. All those works coincide in the idea that support for the molecular $\bar{K} \Xi^*$ state should come from the decay $\Omega^* \rightarrow \bar{K} \pi \Xi$, which actually comes from $\Omega^* \rightarrow \bar{K} \Xi^* (\text{virtual}), \Xi^* \rightarrow \pi \Xi$. In Refs.~\cite{Si-Qi:2016gmh,Huang:2018wth} the evaluation is done for this decay leading to a partial width of about $3$ MeV. In Refs.~\cite{Valderrama:2018bmv,Huang:2018wth} the coupled channels problem is solved to evaluate the couplings of the $\Omega^*$ to $\bar{K} \Xi^*$ and from there the decay width of  $\Omega^* \rightarrow \bar{K} \pi \Xi$ is evaluated. In Ref.~\cite{Lin:2018nqd} the coupling is obtained using the Weinberg compositeness condition \cite{Weinberg:1965zz,Baru:2003qq} and the results for this coupling between \cite{Lin:2018nqd} and \cite{Huang:2018wth} vary by about $20 \%$.

Apart from the $\Omega^* \rightarrow \bar{K} \pi \Xi$ decay channel the $\Omega^* \rightarrow \bar{K} \Xi$ channel is also evaluated in Refs.~\cite{Valderrama:2018bmv,Lin:2018nqd,Huang:2018wth}. In Ref.~\cite{Valderrama:2018bmv} $SU(3)$ arguments are invoked to relate this decay to the $\Delta \rightarrow \pi N$, although the $\Omega^*$ is in a $\frac{3}{2}^-$ state and $\Delta$ in $\frac{3}{2}^+$, which requires d-wave in the $\Omega^*$ case and p-wave in the $\Delta$ case. In Ref.~\cite{Lin:2018nqd} a triangle diagram with $\Omega^* \rightarrow \bar{K} \Xi^*$ exchanging vector mesons is used to make the transition to the final $\bar{K} \Xi$ state. In Ref.~\cite{Huang:2018wth} vector mesons and baryons are allowed to be exchanged in the triangle diagram and the contribution of the vector mesons is found negligible. In Ref.~\cite{Valderrama:2018bmv} the transition is estimated using a naive dimensional analysis from Ref.~\cite{Manohar:1983md}. 

In the present work we follow the lines of Refs.~\cite{Sarkar:2004jh,Valderrama:2018bmv,Lin:2018nqd,Huang:2018wth} and as a novelty we use the $\bar{K} \Xi^*$, $\eta \Omega$ and $\bar{K} \Xi$ channels as coupled channels, with $\bar{K} \Xi^*$ and $\eta \Omega$ in s-wave and $\bar{K} \Xi$ in d-wave. The formalism in this case follows closely the one of Ref.~\cite{Roca:2006sz}. Also, given the fact that some channels are in s-wave and the $\bar{K} \Xi$ in d-wave, and in view of the different subtraction constants required in Ref.~\cite{Roca:2006sz} for those channels, we found more instructive to use cutoff regularization, since the idea of the cutoff is more intuitive, and, as we shall see, one can use the same cutoff in the s- and d-wave channels. One of the outcomes of the full coupled channel is that the decay width for the $\Omega^* \rightarrow \bar{K} \pi \Xi$ is provided directly from the model without the need to study it explicitly. Indeed, the $\Omega^* \rightarrow \bar{K} \pi \Xi$ comes from $\Omega^* \rightarrow \bar{K} \Xi^*$ with the posterior decay $\Xi^* \rightarrow \pi \Xi$. This is incorporated automatically in our scheme by making a convolution of the $\bar{K} \Xi^*$ loop function with the spectral function of the $\Xi^*$ which incorporates the width for the $\pi \Xi$ channel. The other output of the approach is that, given the sensitivity of the model to the input due to the zero diagonal transition matrix elements of the interaction, the inclusion of the $\bar{K} \Xi$ channel into the coupled channels has some effect, producing a shift in the position of the pole (although small) and some diversion in the couplings from the perturbative approach to $\Omega^* \rightarrow\bar{K} \Xi$ done in Refs.~\cite{Lin:2018nqd,Huang:2018wth}. Another new output of the work is the determination of the wave function at the origin for the $\bar{K} \Xi^*$ and $\eta \Omega$ channels that comes to support the dominance of the $\bar{K} \Xi^*$ component in the molecular wave function of the $\Omega^*$. With these differences with respect to the former models, our approach comes to support the conclusions of Refs.~\cite{Valderrama:2018bmv,Lin:2018nqd,Huang:2018wth} as to the natural interpretation of the recent $\Omega (2012)$  state in terms of a dynamically generated resonance from the $\bar{K} \Xi^*$, $\eta \Omega$ and $\bar{K} \Xi$ channels, with the largest overlap to the $\bar{K} \Xi^*$ channel.

\section{Formalism}

The new $\Omega^{*}$ state has been observed mainly in the $\Upsilon(1S)$, $\Upsilon(2S)$ and $\Upsilon(3S)$ decays where the search was made for the decay of $\Omega^{*}$ into $\bar{K}\Xi$. Since the quantum numbers of the $\Omega^{*}$ are more likely to be $J^P=\frac{3}{2}^-$ \cite{Yelton:2018mag}, the coupling of $\bar{K}\Xi$ to $\Omega^{*}$ is, in this case, in d-wave.
And given the spin and flavor structure of the $\Omega^{*}$, it will couple in s-wave to $\bar{K} \Xi^*$ and $\eta \Omega$, and the decay to $\bar{K}\Xi$  will proceed via these channels.

\begin{figure}
\includegraphics[scale=0.5]{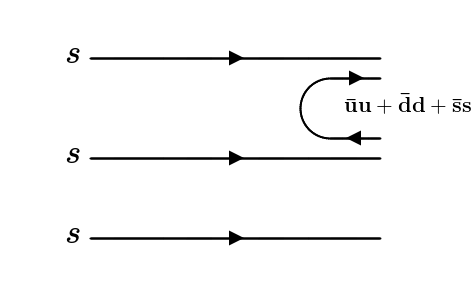}
\centering
\caption{Flavor structure of the $\Omega^{*-}$ decay.}
\label{fig1}
\end{figure}

If we assume that in the decays of $\Upsilon(1S)$, $\Upsilon(2S)$ and $\Upsilon(3S)$ an $sss$ state is formed, then the decay of $\Omega^{*}$  to the available channels will happen as shown in Fig.~\ref{fig1}. For this amplitude to be in s-wave one needs an excited $s$ quark in $L=1$, which is plotted as the upper $s$-quark in the figure.

In terms of the flavor states the hadronization goes as
\begin{equation}
sss \rightarrow \sum_{i=1}^{3} s \bar{q}_i q_i ss \equiv H.
\end{equation}

Then, defining the matrix
\begin{equation}
M=\begin{pmatrix}
u \bar{u} &  u \bar{d} &  u \bar{s}\\ 
d \bar{u} & d \bar{d}& d \bar{s}\\ 
s \bar{u}& s\bar{d} & s \bar{s}
\end{pmatrix},
\end{equation}

we get
\begin{equation}
H= \sum_{i=1}^{3} M_{3i} q_i ss.
\end{equation}
We then rewrite the $q\bar{q}$ matrix in terms of the meson components, $\phi$, as:
\begin{equation}
\phi=\begin{pmatrix}
\frac{\pi^0}{\sqrt{2}} +\frac{\eta}{\sqrt{3}} + \frac{\eta'}{\sqrt{6}} & \pi^+ & K^+\\ 
\pi^- & -\frac{\pi^0}{\sqrt{2}} +\frac{\eta}{\sqrt{3}} + \frac{\eta'}{\sqrt{6}} & K^0\\ 
K^- & \bar{K}^0 &  -\frac{\eta}{\sqrt{3}} + \sqrt{\frac{2}{6}}\eta'
\end{pmatrix},
\end{equation}
where the standard $\eta$, $\eta'$ mixing is assumed \cite{Bramon:1992kr}, and we get
\begin{equation}
\label{eq:flavorH}
H=K^- uss+\bar{K}^0 dss + \left(-\frac{\eta}{\sqrt{3}} + \sqrt{\frac{2}{6}}\eta' \right) sss.
\end{equation}

As usual, we omit the $\eta'$ because of its large mass. 

\begin{figure}
\includegraphics[scale=0.5]{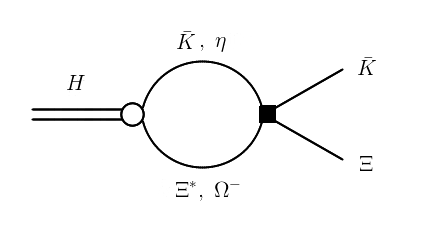}
\centering
\caption{Decay of $H \rightarrow \bar{K} \Xi$ through the creation and re-scattering of the $\bar{K} \Xi^*$ and $\eta \Omega$ pairs.}
\label{fig2}
\end{figure}

We want to write the three quark states in terms of the baryon states which belong to the $SU(3)$ decuplet representation:
\begin{subequations}
\label{eq:xistates}
\begin{align}
& \Xi^{* 0}= \frac{1}{\sqrt{3}} (uss+sus+ssu),\\ 
& \Xi^{*-}= \frac{1}{\sqrt{3}} (dss+sds+ssd), \\
& \Omega = sss.
\end{align}
\end{subequations}
Now, we check the overlap between the baryon states in Eqs.~\eqref{eq:xistates}, and the quark states in Eq.~\eqref{eq:flavorH}:
\begin{subequations}
\begin{align}
& \frac{1}{\sqrt{3}}\left< uss\right|uss+sus+ssu \left. \right> =\frac{1}{\sqrt{3}}, \\
& \frac{1}{\sqrt{3}}\left< dss\right|dss+sds+ssd \left. \right>=\frac{1}{\sqrt{3}}, \\
&  \left< sss\right|sss\left. \right>=1.
\end{align}
\end{subequations}
Finally, we get
\begin{equation}
H=\frac{1}{\sqrt{3}} K^- \Xi^{* 0} + \frac{1}{\sqrt{3}} K^0 \Xi^{*-} - \frac{\eta}{\sqrt{3}} \Omega^-.
\end{equation}
In the isospin basis we have
\begin{equation}
\left|\bar{K} \Xi^*;I=0\right> =\frac{1}{\sqrt{2}} \left( K^- \Xi^{* 0} + K^0 \Xi^{*-} \right),
\end{equation}
where the isospin doublets have the following sign convention: $(\bar{K}^0,-K^-)$ and $(\Xi^{* 0},\Xi^{* -})$.
Hence, the flavor state becomes
\begin{equation}
H=\sqrt{\frac{2}{3}} \left|\bar{K} \Xi^*;I=0\right> - \frac{1}{\sqrt{3}} \eta \Omega^-.
\end{equation}
The transition from $H$ to $\bar{K} \Xi$ will then proceed as shown in Fig.~\ref{fig2}, through the creation and re-scattering of the $\bar{K} \Xi^*$ and $\eta \Omega$ pairs, with the following amplitude:
\begin{widetext}
\begin{equation}
\label{eq:amp}
A(M_{\text{inv}}(\bar{K}\Xi))= \sum_{i=1}^2 h_i G_i(M_{\text{inv}}(\bar{K}\Xi)) t_{i, \bar{K}\Xi}, \ \ \ \text{for } i \equiv \bar{K}\Xi^*, \eta \Omega^-,
\end{equation}
\end{widetext}
where $h_{\bar{K}\Xi^*} = \sqrt{\frac{2}{3}}$; $h_{\eta \Omega^-} = -\frac{1}{\sqrt{3}}$, and $t_{i, \bar{K}\Xi}$ is the amplitude that we will calculate using chiral unitary approach.
In principle one could also have direct $\bar{K}\Xi$ production, but this would just give a background since the $\bar{K}\Xi$ interaction is very weak in $L=2$.

Now, to calculate the $t_{i, \bar{K}\Xi}$ amplitude we can use the work done in Ref.~\cite{Sarkar:2004jh} for the interaction of pseudo-scalar mesons with the baryon decuplet. The potential used there is
\begin{equation}
V_{ij} =-\frac{1}{4f^2} C_{ij} (k^0+k'^0),
\end{equation}
with
\begin{equation}
C=\begin{pmatrix}
0 & 3\\ 
3 & 0
\end{pmatrix},
\end{equation}
and,
\begin{subequations}
\begin{align}
& k^0 = \frac{s+m^2_{\text{in}}-M^2_{\text{in}}}{2\sqrt{2}},\\
& k'^0 = \frac{s+m^2_{\text{fin}}-M^2_{\text{fin}}}{2\sqrt{2}},
\end{align}
\end{subequations}
where $\sqrt{s} \equiv M_{\text{inv}} (\bar{K} \Xi)$.
Here $m_{\text{in}}$, $m_{\text{fin}}$ are the initial and final meson masses and $M_{\text{in}}$, $M_{\text{fin}}$ the initial and final baryon masses.

\begin{figure}
\includegraphics[scale=0.55]{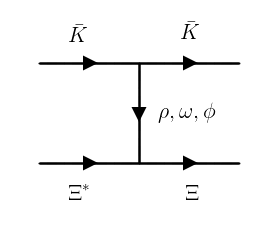}
\centering
\caption{Possible d-wave diagram for the $\bar{K} \Xi^* \rightarrow \bar{K} \Xi$ transition.}
\label{fig3}
\end{figure}

In Ref.~\cite{Sarkar:2004jh} only the s-wave interactions were taken into account and the $m B^* \rightarrow \bar{K} \Xi$ transition is not considered since it is in d-wave. The calculation of this transition from theoretical principles is not easy. If we take the diagram of Fig.~\ref{fig3}, as proposed in Ref.~\cite{Lin:2018nqd}, in the non-relativistic limit one gets an amplitude which is zero, assuming zero initial momentum:
\begin{equation*}
 \left< M \right> \propto \sum_{\epsilon \text{ pol.}} (\vec{\epsilon} \times \vec{q}) \cdot \vec{S} \ \ \vec{\epsilon}\cdot (\vec{p}_{K}+\vec{p}_{K}') = 
\end{equation*}
\begin{equation}
=\sum_{\epsilon \text{ pol.}} (\vec{\epsilon} \times \vec{q}) \cdot \vec{S} \ \ \vec{\epsilon}\cdot \vec{q} = (\vec{q} \times \vec{q})\cdot \vec{S} = 0,
\end{equation}
with $\vec{S}$ the transition operator from $J=\frac{3}{2}$ to $J=\frac{1}{2}$, and $\vec{q}$ the transferred momentum.
This result agrees with the findings of Ref.~\cite{Huang:2018wth}.

\begin{figure}
\includegraphics[scale=0.55]{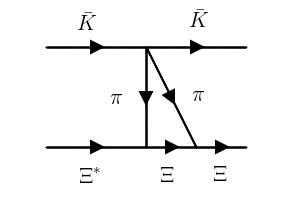}
\centering
\caption{Another possible d-wave diagram for the $\bar{K} \Xi^* \rightarrow \bar{K} \Xi$ transition.}
\label{fig4}
\end{figure}

Then, to get this interaction one would need a structure like the one in Fig.~\ref{fig4}. This corresponds to the $\vec{S}\cdot \vec{q} \ \ \vec{\sigma}\cdot \vec{q}$ structure of Ref.~\cite{Valderrama:2018bmv}. However, the couplings of this diagram are not well known, although in Ref.~\cite{Valderrama:2018bmv} a guess was made regarding the strength of the amplitude. A different approach is done in Ref.~\cite{Huang:2018wth} where the leading terms of the transition come from baryon exchange and the transition is equally subject to large uncertainties. In this work we will not attempt to calculate this diagram. Instead we will leave the strength of the $m B^* \rightarrow \bar{K} \Xi$ transitions as free parameters that will be adjusted by comparing our predictions for the $\Omega^*$ position with experiment. 
We know that these transitions will go as $q^2$ (where $q$ is the momentum of the $\bar{K} \Xi$ channel), then, the three channel interaction matrix will be:
\begin{equation}
\label{eq:Vm}
V= \begin{matrix}
 \begin{matrix}
\bar{K} \Xi^* & \eta \Omega &  \bar{K}\Xi
\end{matrix}& \\ 
\begin{pmatrix}
 0& 3F & \alpha q^2\\ 
 3F& 0 & \beta q^2\\ 
 \alpha q^2 &  \beta q^2 & 0 
\end{pmatrix} & \begin{matrix}
\bar{K} \Xi^* \\ 
\eta \Omega \\ 
\bar{K}\Xi
\end{matrix}
\end{matrix}
\end{equation}
with $F= -\frac{1}{4f^2} (k^0 +k'^0)$, where we assume the interaction of $\bar{K} \Xi \rightarrow \bar{K} \Xi$  in d-wave is very small. Note that the diagonal terms in all the channels are zero. In most cases of molecular states these terms are attractive. One consequence of this feature is the stronger sensitivity to modifications of the parameters in the present case.

Using the potential of Eq.~\ref{eq:Vm} we can then calculate the Bethe-Salpeter equation:
\begin{equation}
\label{eq:BS}
t=\left[1-VG\right]^{-1} V,
\end{equation}
with
\begin{equation}
\label{eq:loopmat}
G(\sqrt{s})=\begin{pmatrix}
G_{\bar{K}\Xi^*}(\sqrt{s}) & 0 & 0\\ 
0 & G_{\eta \Omega}(\sqrt{s}) &0 \\ 
0 & 0 & G_{\bar{K}\Xi}(\sqrt{s})
\end{pmatrix},
\end{equation}
where
\begin{widetext}
\begin{equation}
\label{eq:loop}
G_i(\sqrt{s}) = \int_{|\vec{q}|<q_{\text{max}}} \frac{d^3 q}{(2 \pi)^3} \frac{1}{2 \omega_i(\vec{q})} \frac{M_i}{E_i(\vec{q})} \frac{1}{\sqrt{s}-\omega_i(\vec{q})-E_i(\vec{q}) + i \epsilon} ,
\end{equation}
\end{widetext}
for $i=\bar{K}\Xi^*, \eta \Omega$, with $\omega_i(\vec{q})$ and $E_i(\vec{q})$ the meson and baryon energy, respectively, and $q_{\text{max}}\sim 700$ MeV the cutoff. 
The $G_{\bar{K}\Xi}(\sqrt{s})$ needs to be defined more carefully since the $q$ in Eq.~\eqref{eq:Vm} is the variable of integration of the loop function. For this purpose we substitute, in Eq.~\eqref{eq:Vm}, $q^2$ by $q_{on}^2$, where
\begin{equation}
q_{on} = \frac{\lambda^{1/2}\left(s,m_{K}^2,m_{\Xi}^2,\right)}{2 \sqrt{s}},
\end{equation}
and then the loop function of $\bar{K}\Xi$ becomes:
\begin{widetext}
\begin{equation}
G_{\bar{K}\Xi}(\sqrt{s}) = \int_{|\vec{q}|<q'_{\text{max}}} \frac{d^3 q}{(2 \pi)^3} \frac{ (q/q_{on})^4}{2 \omega_{\bar{K}}(\vec{q})} \frac{M_{\Xi}}{E_{\Xi}(\vec{q})} \frac{1}{\sqrt{s}-\omega_{\bar{K}}(\vec{q})-E_{\Xi}(\vec{q}) + i \epsilon},
\end{equation}
\end{widetext}
where $q'_{\text{max}} \sim 700$ MeV is the cutoff, not necessarily equal to $q_{\text{max}}$.

So far we have four parameters, $\alpha, \ \beta, \ q_{\text{max}} \ \text{and} \  q'_{\text{max}}$. The cutoffs $q_{\text{max}}$ and $q'_{\text{max}}$ are not completely free as they will be varied close to the value ($700$ MeV) proposed in Ref.~\cite{Sarkar:2004jh}. The parameters $\alpha$ and $\beta$ are harder to estimate. In the study of the $\Lambda(1520)$ with the channels $\pi \Sigma^*$, $K \Xi^*$ in s-wave and $\bar{K} N$, $\pi \Sigma$ in d-wave, in Ref.~\cite{Roca:2006sz}, the d-wave parameters for similar interactions were determined to be of the order of $10^{-7}$ MeV$^{-3}$. However, that was using dimensional regularization with a very large subtraction constant ($a\simeq -8$), which makes a comparison with our case not straightforward. 

\begin{figure}
\includegraphics[scale=0.6]{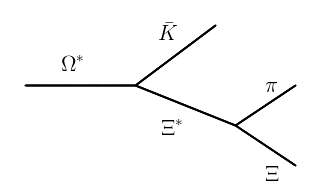}
\centering
\caption{Feynman diagram for the $\Omega^* \rightarrow \bar{K} \pi \Xi$ three body decay.}
\label{fig5}
\end{figure}

In Refs.~\cite{Lin:2018nqd,Huang:2018wth,Si-Qi:2016gmh} the authors claim that most of the width of the $\Omega^*$ comes from the three body decay $\Omega^* \rightarrow \bar{K} \pi \Xi$, with the diagram shown in Fig.~\ref{fig5}. This is the same as considering the two body decay $\Omega^* \rightarrow \bar{K} \Xi^*$ taking into account the mass distribution of $\Xi^*$:
\begin{equation}
S= -\frac{1}{\pi} \text{Im} \left[\frac{1}{M_{\text{inv}}(\Xi^*) - M_{\Xi^*} + i \frac{\Gamma_{\Xi^*}}{2}}\right]=\\
\frac{1}{\pi} \frac{\Gamma_{\Xi^*}/2}{\left(M_{\text{inv}}(\Xi^*) - M_{\Xi^*}\right)^2+\left(\frac{\Gamma_{\Xi^*}}{2}\right)^2}
\end{equation}
Since the $\bar{K} \Xi^*$ threshold is very close to the $\Omega^*$ position, considering the mass distribution of $\Xi^*$ is important. Technically this is accomplished by substituting $G_{\bar{K} \Xi^*}$ in Eqs.~\eqref{eq:amp} and~\eqref{eq:loopmat} by
\begin{widetext}
\begin{equation}
\label{eq:loopconv}
\tilde{G}_{\bar{K} \Xi^*} (\sqrt{s}) = \frac{1}{N} \int_{M_{\Xi^*}-\Delta M_{\Xi^*}}^{M_{\Xi^*}+\Delta M_{\Xi^*}} d \tilde{M} \left(-\frac{1}{\pi} \right) \text{Im} \left(\frac{1}{\tilde{M}-M_{\Xi^*}+i \frac{\Gamma_{\Xi^*}}{2}} \right) \ G_{\bar{K} \Xi^*} (\sqrt{s},m_{\bar{K}},\tilde{M}),
\end{equation} 
with,
\begin{equation}
N= \int_{M_{\Xi^*}-\Delta M_{\Xi^*}}^{M_{\Xi^*}+\Delta M_{\Xi^*}}  d \tilde{M} \left(-\frac{1}{\pi} \right) \text{Im} \left(\frac{1}{\tilde{M}-M_{\Xi^*}+i \frac{\Gamma_{\Xi^*}}{2}} \right),
\end{equation}
\end{widetext}
and we choose $\Delta M_{\Xi^*}= 6 \Gamma_{\Xi^*}$, which takes into account most of the distribution.
It is worth noting that for small $\Gamma_{\Xi^*} \rightarrow 2\epsilon$, we have that
\begin{equation*}
S=\left(-\frac{1}{\pi} \right) \text{Im} \left(\frac{1}{\tilde{M}-M_{\Xi^*}+i \epsilon} \right)
\end{equation*}
\begin{equation}
= \delta(\tilde{M}-M_{\Xi^*}),
\end{equation}
and we recover the original loop function, $ \tilde{G}_{\bar{K} \Xi^*} (\sqrt{s}) = G_{\bar{K} \Xi^*} (\sqrt{s})$.

\section{Results}

In Ref.~\cite{Yelton:2018mag}, the $\Omega^*$ was measured to have
\begin{subequations}
\begin{align}
& m_{\Omega^*}^{\text{exp}}= 2012.4 \pm 0.92 \text{ MeV}, \\
& \Gamma_{\Omega^*}^{\text{exp}} = 6.4^{+3.0}_{-2.6} \text{ MeV}.
\end{align}
\end{subequations}
Then, by calculating the position of the poles in the $t$ matrix of Eq.~\eqref{eq:BS}, we can determine the best set of parameters that can reproduce the experimental results. The search will be conducted by going to the second Riemann sheet (SRS) above the thresholds of the different channels, which corresponds to using the following loop functions:
\begin{equation}
G_{\text{SRS}}^i (\sqrt{s}) = G_i(\sqrt{s}) + \left\{\begin{matrix}
0, &  \text{for } \text{Re}(\sqrt{s})<\sqrt{s}_{th}\\ 
i\frac{M_i k}{2 \pi \sqrt{s}}, &  \text{for } \text{Re}(\sqrt{s})>\sqrt{s}_{th}
\end{matrix}\right. ,
\end{equation}
with
\begin{equation}
k = \frac{\lambda^{1/2}\left(s,m_i^2,M_i^2 \right)}{2 \sqrt{s}}.
\end{equation}
A good set of parameters that fits our conditions is shown in Tab.~\ref{tab1}, and they produce a pole at:
\begin{subequations}
\begin{align}
\label{eq:massfit}
& m_{\Omega^*}= 2012.37 \text{ MeV}, \\
\label{eq:widthfit}
& \Gamma_{\Omega^*} = 6.24 \text{ MeV}.
\end{align}
\end{subequations}

\begin{table}[h]
\centering
\begin{tabular}{ c|c|c } 
 $\alpha$ (MeV$^{-3}$) &  $\beta$ (MeV$^{-3}$)  & $q_{\text{max}}=q'_{\text{max}}$ (MeV) \\
 \hline
  $4.0 \times 10^{-8}$ & $1.5 \times 10^{-8}$ & 735 
\end{tabular}
\caption{Parameters from a fit to the experimental results.}
\label{tab1}
\end{table}

The couplings of the channels to the resonance can be calculated using
\begin{equation}
t_{ij} = \frac{g_i g_j}{\sqrt{s}-M_R+i \frac{\Gamma_R}{2}},
\end{equation}
and the results are presented in Table.~\ref{tab2}. The couplings $g_{\bar{K} \Xi^*}$ and $g_{\eta \Omega}$ obtained here are close to the ones of Ref.~\cite{Huang:2018wth} and also to the $g_{\bar{K} \Xi^*}$ from Ref.~\cite{Lin:2018nqd}.

One can also compare the order of magnitude of our $V_{i,\bar{K} \Xi}$ with the one estimated in Ref.~\cite{Valderrama:2018bmv}. There
\begin{equation}
 \left< \bar{K} \Xi^*\right| \hat{V} \left| \bar{K} \Xi \right> = C_D \ \vec{\sigma}\cdot \vec{q}  \ \ \vec{S}\cdot \vec{q},
\end{equation}
where
\begin{equation}
C_D \sim \frac{1}{f^2 \Lambda_{\chi }},
\end{equation}
with $\Lambda_{\chi }=1000$ MeV.
Then, we can make use of the following relation
\begin{equation}
\sum_{M_s}  S_k \left|M_s \right> \left< M_s\right| S_l^{\dagger} =\frac{2}{3}\delta_{kl} - \frac{i}{3} \epsilon_{klm}\sigma_m,
\end{equation} 
to calculate the square of the interaction
\begin{equation*}
\sum \overline{\sum} \left| \left< \bar{K} \Xi^*\right| \hat{V} \left| \bar{K} \Xi \right>\right|^2=V^2_{\bar{K} \Xi^*,\bar{K} \Xi} = C_D^2 \overline{\sum}\sum_{m_s} \left< m_s\right| \sigma_i  \left(\frac{2}{3}\delta_{kl} - \frac{i}{3} \epsilon_{klm}\sigma_m \right)  \sigma_j^{\dagger} \left|m_s \right> \ q_i q_j q_k q_l
\end{equation*}
\begin{equation}
= \frac{1}{4} \sum_{m_s} \frac{2}{3} C_D^2 \left< m_s\right| q^4   \left|m_s \right> = \frac{1}{3} q^4 C_D^2 = \alpha^2 q^4. 
\end{equation}
Then, if $f=f_{\pi}=93 \text{ MeV}(f_K=160 \text{ MeV})$, we will get $C_D= 1.2 \times 10^{-7} \text{ MeV}^{-3} (3.9 \times 10^{-8}\text{ MeV}^{-3})$, which gives $\alpha= 4.7 \times 10^{-8} \text{ MeV}^{-3} (1.6 \times 10^{-8}\text{ MeV}^{-3})$. This means that our parameters $\alpha, \ \beta$ agree, at least in the order of magnitude, with Ref.~\cite{Valderrama:2018bmv}.

We can also check the contribution of the three particle decay channel in Fig.~\ref{fig5} by calculating the pole position without the convolution:
\begin{subequations}
\begin{align}
\label{qeq:conv1}
& m_{\Omega^*}^{\text{(no conv.)}}= 2013.5 \text{ MeV}, \\
\label{qeq:conv2}
& \Gamma_{\Omega^*}^{\text{(no conv.)}} = 3.2 \text{ MeV}.
\end{align}
\end{subequations}
Then the difference of the widths $\Gamma_{\Omega^*}  - \Gamma_{\Omega^*}^{\text{(no conv.)}} = 3$ MeV can be attributed to the $\Omega^* \rightarrow \bar{K} \Xi^* \rightarrow \bar{K} \pi \Xi$ decay, which is similar to the three body contribution found in Refs.~\cite{Huang:2018wth,Lin:2018nqd}. The remaining $3$ MeV would correspond to the $\bar{K}\Xi$ decay channel.

In Ref.~\cite{Huang:2018wth} a state very near $\Omega^*$ is obtained for a subtraction constant of $a=-2.5$, which is equivalent to using a cutoff of about $\Lambda \simeq 730$ MeV, which is very close to our value in Tab.~\ref{tab1}. Furthermore, we can test the effect of the cutoff changing it by $100$ MeV ($q_{\text{max}}=835$ MeV), while keeping the other parameters fixed. Then, the pole position shifts to:
\begin{subequations}
\begin{align}
& m_{\Omega^*}= 1982.23 \text{ MeV}, \\
& \Gamma_{\Omega^*} = 3.13 \text{ MeV}.
\end{align}
\end{subequations}
Note that, because the distance to $\bar{K} \Xi^*$ is now bigger, this decay channel will have a smaller strength. However, there will still be an effect from the convolution. This can be seen by removing the convolution and we find that the new width is $2.94$ MeV, which means that $0.19$ MeV comes from the $\bar{K} \Xi^*$ ($\bar{K} \pi \Xi$) decay and $2.94$ MeV from the $\bar{K} \Xi$ decay. We can see that the partial decay width into $\bar{K} \Xi$ is rather independent of the cutoff.
\begin{table}[h]
\centering
\begin{tabular}{ c|c|c}
\hspace{0.2cm} $g_{\bar{K} \Xi^*}$ \hspace{0.2cm}    &  \hspace{0.2cm}  $g_{\eta \Omega}$ \hspace{0.2cm} & \hspace{0.2cm} $g_{\bar{K} \Xi}$ \hspace{0.2cm}  \\ \hline
$2.01+i0.02$ & $2.84-i0.01$ & $-0.29+i0.04$
\end{tabular}
\caption{Couplings of the $\Omega^*$ to the three channels.}
\label{tab2}
\end{table}

\begin{figure}
\includegraphics[scale=0.45]{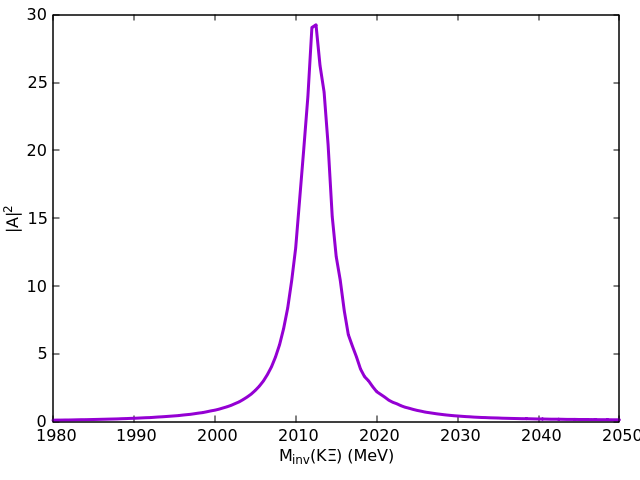}
\centering
\caption{Plot of the modulus square of the amplitude $A$ of $H \rightarrow \bar{K} \Xi$ (see Eq.~\eqref{eq:amp}) as a function of $M_{\text{inv}}(\bar{K} \Xi)$.}
\label{fig6}
\end{figure}

\begin{figure}
\includegraphics[scale=0.5]{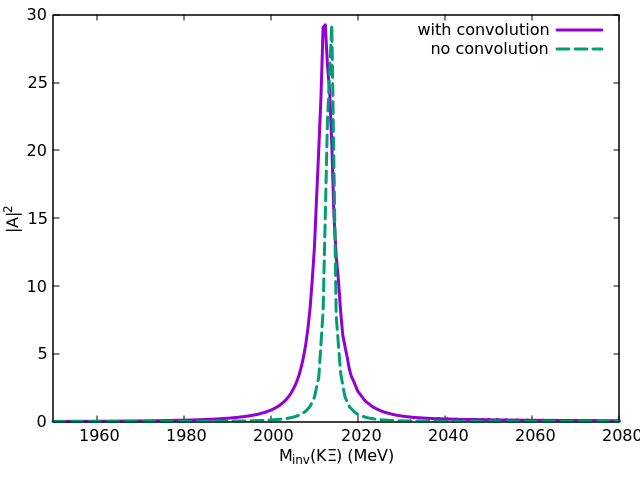}
\centering
\caption{Plot of $|A|^2$ calculated using convolution (full line) and by not using the convolution (dashed line) normalized to the peak.}
\label{fig8}
\end{figure}

Finally, the amplitude for the $H \rightarrow \bar{K} \Xi$ can be obtained using Eq.~\eqref{eq:amp} with the loop function of Eq.~\eqref{eq:loopconv} for the  $\bar{K} \Xi^*$ channel, and we obtain the curve shown in Fig.~\ref{fig6}. It is worth noting that the $|T_{\eta \Omega \rightarrow \eta \Omega}|^2$ amplitude has a shape indistinguishable from the one shown in Fig.~\ref{fig6}. In Fig.~\ref{fig8} we also compare $|A|$ from Eq.~\eqref{eq:amp} with what we would obtain without the convolution. One can see that both the energy position and the width of the $\Omega^*$ state change according to the pole position and widths found in Eqs.~\eqref{qeq:conv1} and~\eqref{qeq:conv2}.

Another magnitude that stresses the relevance of the different channels is the wave function at the origin, given by $\left.-g^2 \frac{\partial G(\sqrt{s})}{\partial \sqrt{s}}\right|_{s=s_R}$~\cite{Gamermann:2009uq}. The results are shown in Tab.~\ref{tab3} for the two s-wave channels. This information is relevant because, even if the coupling of the state to $\eta \Omega$ is $1.4$ times bigger than to $\bar{K} \Xi^*$, the wave function at the origin for $\bar{K} \Xi^*$ is four times bigger than the one of $\eta \Omega$, as a consequence of being closer in energy to the resonance.

\begin{table}[h]
\centering
\begin{tabular}{ c|c} 
\hspace{0.2cm}  $\left(-g^2 \frac{\partial G(\sqrt{s})}{\partial \sqrt{s}}\right)_{\bar{K} \Xi^*}$ \hspace{0.2cm} & \hspace{0.2cm} $\left(-g^2 \frac{\partial G(\sqrt{s})}{\partial \sqrt{s}}\right)_{\eta \Omega}$ \hspace{0.2cm} \\
 \hline 
\hspace{0.2cm}  $0.636-i 0.068$ \hspace{0.2cm} &\hspace{0.2cm} $0.164-i 0.002$\hspace{0.2cm}
\end{tabular}
\caption{Values of the wave function at the origin for the two s-wave channels.}
\label{tab3}
\end{table}

\begin{figure}
\includegraphics[scale=0.55]{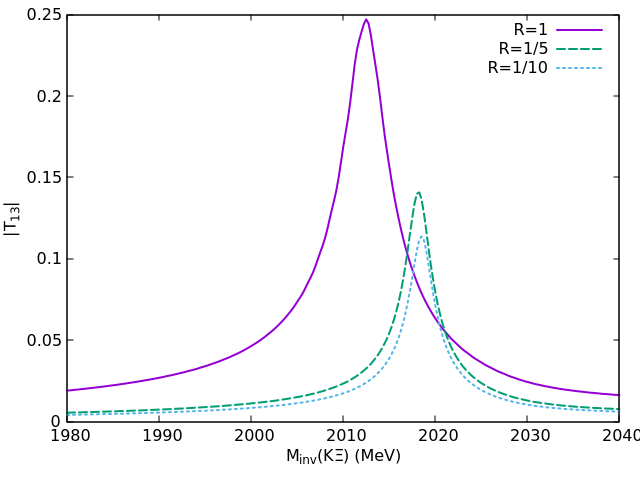}
\centering
\caption{Comparing $|T_{13}|$ using $R\alpha$ as a parameter, for different values of R.}
\label{fig9}
\end{figure}

We can also study the importance of the $\bar{K} \Xi$ channel by multiplying the $\alpha$ parameter by a factor $R$. In Fig.~\ref{fig9} we compare the behavior of the amplitude $|T_{13}|$ for $R=1,\frac{1}{5}, \frac{1}{10}$. Two interesting things can be seen there, first, $|T_{13}(R)| \neq R |T_{13}|$, or in other words, multiplying $\alpha$ by a factor $R$ is not equivalent to multiplying the amplitude by the same factor. Also, by changing $R$ to lower values, there is a shift in the pole position of about $6$ MeV. This means that the addition of the $\bar{K} \Xi$ channel is important. The stability of the results of Fig.~\ref{fig9} when $R \rightarrow 0$ is telling us that about half of the strength in $T_{13}$ comes from the intermediate $\eta \Omega$ channel. This is worth mentioning since when working with coupled $\bar{K} \Xi^*$ and $\eta \Omega$ channels, this is implemented automatically, as done here and in~\cite{Huang:2018wth}. Yet, if one takes the $\bar{K} \Xi^*$ dominant channel alone, as done in~\cite{Lin:2018nqd}, this contribution would be vanishing.

In Ref.~\cite{Huang:2018wth} it was found that the transitions from $\bar{K} \Xi^*$ and $\eta \Omega$ to $\bar{K} \Xi$ depended strongly on one parameter. Yet the model used there gave a very small contribution of the $\eta \Omega \rightarrow \bar{K} \Xi$ transition on its own, although when summed coherently to the one of $\bar{K} \Xi^*$ it was noticeable. We would like to investigate if some other fit in our approach, with a negligible $\eta \Omega \rightarrow \bar{K}\Xi$ transition is possible. For this purpose we decrease $\beta$ in Eq.~\eqref{eq:Vm} by a factor 10. We can see that a reasonable fit to the mass and width is also possible. In this case we obtain the parameters of Table~\ref{tab4}. We obtain now the results:
\begin{subequations}
\begin{align}
\label{eq:small1}
& m_{\Omega^*}= 2012.09 \text{ MeV}, \\
\label{eq:small2}
& \Gamma_{\Omega^*} = 6.41 \text{ MeV},
\end{align}
\end{subequations}
which are very similar to these in Eqs.~\eqref{eq:massfit} and~\eqref{eq:widthfit}. If we perform the calculation without the convolution we obtain
\begin{subequations}
\begin{align}
& m_{\Omega^*}= 2011.8 \text{ MeV}, \\
& \Gamma_{\Omega^*} = 4.07 \text{ MeV},
\end{align}
\end{subequations}
which differ a bit with respect to those in Eqs.~\eqref{qeq:conv1} and~\eqref{qeq:conv2}. This means that now the contribution of $\Omega^* \rightarrow \bar{K}\Xi$ would be about $4$ MeV, somewhat bigger than in Eq.~\eqref{qeq:conv2}. An explicit measurement of the $\Omega^*$ partial decay widths to $\bar{K} \Xi^* \rightarrow \bar{K} \Xi \pi$ and $\bar{K}\Xi$ would provide further information to settle these present theoretical uncertainties.

\begin{table}[h]
\centering
\begin{tabular}{ c|c|c } 
 $\alpha$ (MeV$^{-3}$) &  $\beta$ (MeV$^{-3}$)  & $q_{\text{max}}=q'_{\text{max}}$ (MeV) \\
 \hline
  $5.0 \times 10^{-8}$ & $1.5 \times 10^{-9}$ & 735 
\end{tabular}
\caption{Parameters from the fit to the experimental results with small $\beta$.}
\label{tab4}
\end{table}

\section{Conclusions}

The recent observation of the excited $\Omega^* (2012)$ state has brought the necessary experimental information to complete the chiral unitary approach that generates resonances from the interaction of pseudo-scalar  mesons and baryons of the $\Delta$ decuplet. We have performed the coupled channels problem using the $\bar{K} \Xi^*$, $\eta \Omega$ and $\bar{K} \Xi$ states, the first two channels in s-wave and the latter one in d-wave. The incorporation of the $\bar{K} \Xi$ channel in the set of coupled channels is a novelty of our approach, and although not too strong, we see that it has some effects on the mass of the state and couplings that go beyond the perturbative treatment of this channel. We have also shown that the $\Omega^* \rightarrow \bar{K} \pi \Xi$ decay is relevant, coinciding with several recent studies, but its evaluation is done differently since we see that this decay is a direct output of the coupled channels as soon as the spectral function of the $\Xi^*$ is used to evaluate the $\bar{K} \Xi^*$ loop function.

Finally we have also shown that, apart of the coupling of the $\Omega^*$ to the different channels, the wave function at the origin is important, and looking at this magnitude one can see that the $\bar{K} \Xi^*$ component is largely dominant in the $\Omega^*$ molecular wave function. This makes this molecule peculiar since in the absence of the $\eta \Omega$ channel the $\bar{K} \Xi^*$ system does not bind. The introduction of the $\eta \Omega$ channel produces a bound state, which couples more strongly to the $\bar{K} \Xi^*$ channel due to the proximity of the $\Omega^*$ mass to the $\bar{K} \Xi^*$ threshold.

With the novelties introduced in our approach we come to support former findings indicating that the new $\Omega^* (2012)$ state stands naturally for a molecular state with $\bar{K} \Xi^*$ as its main component.

\section*{Acknowledgements}
One of us, R.P, wishes to acknowledge the Generalitat Valenciana in the program Santiago Grisolia. This work is partly supported
by the Spanish Ministerio de Economia y Competitividad and European FEDER funds
under Contracts No. FIS2017-84038-C2-1-P B and No. FIS2017-84038-C2-2-P B, and the
Generalitat Valenciana in the program Prometeo II-2014/068, and the project Severo Ochoa
of IFIC, SEV-2014-0398.

\bibliographystyle{plain}

\end{document}